\documentclass[aps,prc,superscriptaddress,showpacs,amssymb,amsmath,amsfonts,twocolumn,floatfix]{revtex4}

\usepackage{graphicx}
\usepackage{dcolumn}
\usepackage{bm}

\begin{document}


\title{Beam spin asymmetry in deep and exclusive $\pi^{0}$
	electroproduction 
	}

\newcommand*{\SACLAY}{CEA-Saclay, Service de Physique Nucl\'eaire, 91191 Gif-sur-Yvette, France}
\affiliation{\SACLAY}
\newcommand*{\UCONN}{University of Connecticut, Storrs, Connecticut 06269}
\affiliation{\UCONN}

\newcommand*{\ANL}{Argonne National Laboratory, Argonne, Illinois 60439}
\affiliation{\ANL}
\newcommand*{\ASU}{Arizona State University, Tempe, Arizona 85287-1504}
\affiliation{\ASU}
\newcommand*{\UCLA}{University of California at Los Angeles, Los Angeles, California  90095-1547}
\affiliation{\UCLA}
\newcommand*{\CSU}{California State University, Dominguez Hills, Carson, CA 90747}
\affiliation{\CSU}
\newcommand*{\CMU}{Carnegie Mellon University, Pittsburgh, Pennsylvania 15213}
\affiliation{\CMU}
\newcommand*{\CUA}{Catholic University of America, Washington, D.C. 20064}
\affiliation{\CUA}
\newcommand*{\CNU}{Christopher Newport University, Newport News, Virginia 23606}
\affiliation{\CNU}
\newcommand*{\ECOSSEE}{Edinburgh University, Edinburgh EH9 3JZ, United Kingdom}
\affiliation{\ECOSSEE}
\newcommand*{\FU}{Fairfield University, Fairfield CT 06824}
\affiliation{\FU}
\newcommand*{\FIU}{Florida International University, Miami, Florida 33199}
\affiliation{\FIU}
\newcommand*{\FSU}{Florida State University, Tallahassee, Florida 32306}
\affiliation{\FSU}
\newcommand*{\GWU}{The George Washington University, Washington, DC 20052}
\affiliation{\GWU}
\newcommand*{\ECOSSEG}{University of Glasgow, Glasgow G12 8QQ, United Kingdom}
\affiliation{\ECOSSEG}
\newcommand*{\ISU}{Idaho State University, Pocatello, Idaho 83209}
\affiliation{\ISU}
\newcommand*{\INFNFR}{INFN, Laboratori Nazionali di Frascati, 00044 Frascati, Italy}
\affiliation{\INFNFR}
\newcommand*{\INFNGE}{INFN, Sezione di Genova, 16146 Genova, Italy}
\affiliation{\INFNGE}
\newcommand*{\ITEP}{Institute of Theoretical and Experimental Physics, Moscow, 117259, Russia}
\affiliation{\ITEP}
\newcommand*{\LPC}{LPC Clermont-Ferrand, Universit\'e Blaise Pascal, CNRS/IN2P3, 63177 Aubi\`ere, France}
\affiliation{\LPC}
\newcommand*{\LPSC}{LPSC, Universit\'e Joseph Fourier, CNRS/IN2P3, INPG, 38026 Grenoble, France}
\affiliation{\LPSC}
\newcommand*{\JMU}{James Madison University, Harrisonburg, Virginia 22807}
\affiliation{\JMU}
\newcommand*{\KYUNGPOOK}{Kyungpook National University, Daegu 702-701, South Korea}
\affiliation{\KYUNGPOOK}
\newcommand*{\UMASS}{University of Massachusetts, Amherst, Massachusetts  01003}
\affiliation{\UMASS}
\newcommand*{\MOSCOW}{Moscow State University, General Nuclear Physics Institute, 119899 Moscow, Russia}
\affiliation{\MOSCOW}
\newcommand*{\UNH}{University of New Hampshire, Durham, New Hampshire 03824-3568}
\affiliation{\UNH}
\newcommand*{\NSU}{Norfolk State University, Norfolk, Virginia 23504}
\affiliation{\NSU}
\newcommand*{\OHIOU}{Ohio University, Athens, Ohio  45701}
\affiliation{\OHIOU}
\newcommand*{\ODU}{Old Dominion University, Norfolk, Virginia 23529}
\affiliation{\ODU}
\newcommand*{\ORSAY}{Institut de Physique Nucl\'eaire, 91406 Orsay, France}
\affiliation{\ORSAY}
\newcommand*{\RPI}{Rensselaer Polytechnic Institute, Troy, New York 12180-3590}
\affiliation{\RPI}
\newcommand*{\RICE}{Rice University, Houston, Texas 77005-1892}
\affiliation{\RICE}
\newcommand*{\URICH}{University of Richmond, Richmond, Virginia 23173}
\affiliation{\URICH}
\newcommand*{\SCAROLINA}{University of South Carolina, Columbia, South Carolina 29208}
\affiliation{\SCAROLINA}
\newcommand*{\JLAB}{Thomas Jefferson National Accelerator Facility, Newport News, Virginia 23606}
\affiliation{\JLAB}
\newcommand*{\UNIONC}{Union College, Schenectady, NY 12308}
\affiliation{\UNIONC}
\newcommand*{\VT}{Virginia Polytechnic Institute and State University, Blacksburg, Virginia   24061-0435}
\affiliation{\VT}
\newcommand*{\VIRGINIA}{University of Virginia, Charlottesville, Virginia 22901}
\affiliation{\VIRGINIA}
\newcommand*{\WM}{College of William and Mary, Williamsburg, Virginia 23187-8795}
\affiliation{\WM}
\newcommand*{\YEREVAN}{Yerevan Physics Institute, 375036 Yerevan, Armenia}
\affiliation{\YEREVAN}
\newcommand*{\NOWUNH}{University of New Hampshire, Durham, New Hampshire 03824-3568}
\newcommand*{\NOWUMASS}{University of Massachusetts, Amherst, Massachusetts  01003}
\newcommand*{\NOWMIT}{Massachusetts Institute of Technology, Cambridge, Massachusetts  02139-4307}
\newcommand*{\NOWECOSSEE}{Edinburgh University, Edinburgh EH9 3JZ, United Kingdom}
\newcommand*{\NOWGEISSEN}{Physikalisches Institut der Universitaet Giessen, 35392 Giessen, Germany}

\author {R.~De~Masi} 
\affiliation{\SACLAY}
\affiliation{\ORSAY}
\author {M.~Gar\c con} 
\affiliation{\SACLAY}
\author {B.~Zhao} 
\affiliation{\UCONN}

\author {M.J.~Amaryan} 
\affiliation{\ODU}
\author {P.~Ambrozewicz} 
\affiliation{\FIU}
\author {M.~Anghinolfi} 
\affiliation{\INFNGE}
\author {G.~Asryan} 
\affiliation{\YEREVAN}
\author {H.~Avakian} 
\affiliation{\JLAB}
\author {H.~Bagdasaryan} 
\affiliation{\ODU}
\author {N.~Baillie} 
\affiliation{\WM}
% ...

\collaboration{The CLAS Collaboration}
     \noaffiliation

\date{\today}

\begin{abstract}
The beam spin asymmetry (BSA) in the exclusive reaction $\vec{e}p\rightarrow
ep\pi^0$ was measured with the CEBAF 5.77~GeV polarized electron beam 
and Large Acceptance Spectrometer
(CLAS).   
      The $x_B$, $Q^2$, $t$ and $\phi$ dependences of the
$\pi^0$ BSA are presented in the deep inelastic regime.  The
asymmetries are fitted with a $\sin \phi$ function and their amplitudes are
extracted. Overall, they are of the order of 0.04 - 0.11 and 
roughly independent of $t$. This is the signature of a non-zero
longitudinal-transverse 
interference.
The implications concerning the applicability
of a formalism based on generalized parton distributions,
as well as the extension of a Regge formalism at
high photon virtualities, are discussed.
\end{abstract}

\maketitle

\section{Introduction}
\label{sec:intro}
\begin{figure}
 	\includegraphics[width=5cm]{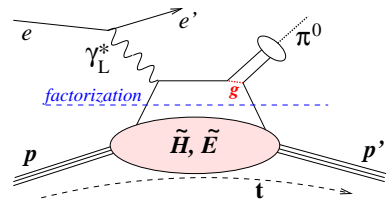}
 \caption{\label{fig:handbag}
 		(Color online)
  	   Schematic representation of the handbag diagram for 
 	   neutral pion production. 
	   The symbol $g$ stands for a gluon exchange between quark lines.
 	 }
\end{figure}

Deeply virtual exclusive reactions $\gamma^*N\to N \gamma, N \pi, N \rho\cdots$,
where the $\gamma^*$ virtuality $Q^2$ is large, 
have the potential to probe nucleon structure
at the parton level, as described by generalized parton distributions
(GPDs). These distributions are universal functions which parameterize 
the non-perturbative structure of the nucleon. They include as limiting cases 
form factors and parton
distributions, and they also provide access to hitherto unknown observables
like the spatial distribution of partons of given longitudinal
momentum fraction or the angular momentum of quarks and
gluons inside the nucleon~\cite{Ji97,Bur00,Die03}. 
The description of deeply virtual meson production in terms
of GPDs relies on a factorization theorem~\cite{Col97}, 
which applies when the virtual photon $\gamma^*$  
is longitudinally polarized. In other words, meson
production is expected to proceed mostly through longitudinal
virtual photons in
the Bjorken regime ($Q^2\rightarrow \infty$ and 
the Bjorken variable $x_{B}$ finite). 
The corresponding leading-twist diagram 
(or handbag diagram, illustrated in Fig.~\ref{fig:handbag}) 
for $\pi^0$ production is sensitive
to specific flavour combinations of quark-helicity dependent (or
``polarized'') GPDs: $\frac{2}{3}\tilde{H}^u+\frac{1}{3}\tilde{H}^d$ and
$\frac{2}{3}\tilde{E}^u+\frac{1}{3}\tilde{E}^d$~\cite{Die03}. 
The $\tilde{H}^q$ are partly constrained by the polarized parton
distributions $\Delta q$, while the $\tilde{E}^q$, largely
unknown, are often modeled by a pion-pole term, 
which would not contribute to the $ep\to ep\pi^0$ 
process~\cite{Die03}.
The $Q^2$ range in which the handbag diagram dominates,
or where its contribution can be safely extracted, is not yet known
for meson production.

An alternative description of exclusive meson production is based on
Regge models, where trajectories are exchanged in the $t$-channel as
mediators of the interaction. While 
extensively studied for photoproduction~\cite{Gui97}, i.e. for $Q^2=0$ and transverse
photons, the extension and applicability to virtual photons
has not yet been considered in the specific case of neutral pion production.

So, two theoretical descriptions are {\sl a priori} possible.
The Regge approach starts from $Q^2=0$ and must be extended
to non-zero $Q^2$, while the GPD approach has a firm QCD foundation
in the Bjorken regime, and its applicability must be tested at
finite values of $Q^2$.

On the experimental side, while the focus has recently been on the
production of real photons~\cite{Gar06} (the so-called deeply virtual
Compton scattering process, or DVCS) and of vector
mesons~\cite{Air00,Had05,Mor05}, there is essentially no experimental
data available on neutral pseudoscalar meson production 
above the resonance region.  
Cross sections were measured at DESY~\cite{Bra75} at low values of $Q^2$, 
while a first result on the target-spin asymmetry was obtained at CLAS~\cite{Che06}. 
For recent data on charged pion electroproduction in this kinematic regime, 
see Refs.~\cite{Air07,Hor07}.

The $ep\to ep \pi^0$ observables depend  on the $Q^2$ and $x_B$ variables, 
                    on the squared four-momentum transfer $t$
to the proton and on the angle $\phi$ between the leptonic
and hadronic planes. 
The polarization of the exchanged virtual photon
may be transverse ($T$) or longitudinal ($L$). 
It induces an azimuthal dependence of the reduced cross section for the
$\gamma^*p\to p \pi^0 $ process. For each ($x_{B}$, $Q^2$, $t$), 
taking the ratio of the difference over the sum of cross sections for opposite
beam helicities, the beam spin asymmetry (BSA) has the following $\phi$ 
dependence:
\begin{equation}
\label{eq:bsa}
A = 
	\frac{\overrightarrow{\sigma}-\overleftarrow{\sigma}}
       		{\overrightarrow{\sigma}+\overleftarrow{\sigma}}=
	\frac{\alpha \sin\phi}{1+\beta\cos\phi+\gamma\cos 2\phi}.
\end{equation} 
\noindent The parameter $\alpha$ is proportional to a term 
denoted $\sigma_{LT^\prime}$, originating from the imaginary part
of an interference between the helicity amplitudes describing the process~\cite{Dre92}.
\begin{equation}
\label{eq:alpha}
\alpha = \frac{\sqrt{2\epsilon(1-\epsilon)}\sigma_{LT^\prime}}
              {\sigma_T+\epsilon \sigma_L} ,
\end{equation}
\noindent where $\sigma_T$ and $\sigma_L$ are the pure transverse and longitudinal 
cross sections, and $\epsilon$ is the usual virtual photon polarization
parameter.
Any measurement of a non-zero BSA would be indicative of an $L$-$T$
interference, and therefore of contributions  
that cannot be described in terms of GPDs.
Indeed, preliminary CLAS data indicated sizeable BSA
both for exclusive $\pi^+$ and $\pi^0$ production at large $Q^2$~\cite{Ava03}.

\section{Experiment and data analysis}
\label{sec:exp}
This experiment used the CEBAF 5.77~GeV longitudinally-polarized electron
beam impinging on a 2.5-cm long liquid-hydrogen target.  
The beam helicity was switched pseudo-randomly at a frequency of 30 Hz,
and the beam polarization, measured with a M{\o}ller polarimeter,
had an average value of 79.4\%.
All final-state particles from the reaction $ep\to ep\pi^0$
followed by the decay $\pi^0\to\gamma\gamma$ were detected.
The six-sector CLAS spectrometer~\cite{Mec03}
was used in order to detect scattered electrons, recoil protons and
photons emitted at large angles.
An additional small electromagnetic calorimeter 
ensured photon detection in the
near forward region (4.5 - 15$^\circ$). 
This inner calorimeter (IC) was built of 424 tapered lead-tungstate crystals,
read out with avalanche photodiodes.
It was calibrated 
using the two-photon decay of (inclusively produced) neutral pions.

Events were selected if an electron had generated a trigger,
one and only one proton was identified and 
any number of photons (above an energy threshold of 150 MeV) were detected in
either the IC or the standard CLAS calorimeter EC~\cite{Ama01}. 
Electrons were identified
through signals in the EC and in the \v{C}erenkov counters.
Events considered hereafter included the kinematic requirements~:
$Q^2>1$~GeV$^2$, $\gamma^*p$ invariant mass $W>2$~GeV and 
scattered electron energy $E'>0.8$~GeV. 
Protons were unambiguously
identified
over the whole momentum range of interest
using time-of-flight from the target to the CLAS scintillators, 
as well as the track length and momentum determined by the drift chambers.
A cut at $\pm 3\sigma$ was applied around the pion mass
in the squared missing mass MM$^2(ep\to epX)$ distribution 
to exclude multipion background.

All clusters detected in the IC were assumed to originate from photons,
while additional time-of-flight information was used in the EC to separate
photons from neutrons. Photons hitting the calorimeters' edges were excluded.
In addition, since the most forward hits in the IC
had a sizeable probability of originating from M{\o}ller accidental
coincidences, 
a minimal angle was imposed on all photon candidates:
$\theta_{\gamma} > 8^{\circ} - 0.75^{\circ} \times (E_{\gamma}/1\text{ GeV})$.

In order to reconstruct the $\pi^0$ candidates, 
the two most energetic detected photons were considered,
originating from either calorimeter.
Four combinations are then possible: IC-IC, IC-EC, EC-IC, EC-EC, where the
photon with the highest energy was in the first mentioned
calorimeter.
The two calorimeters (IC and EC) had similar angular resolutions
(about 4 mrad for 1 GeV photons) but different
energy resolutions 
($\sigma_{E_{\gamma}}/E_{\gamma}\simeq$ 4.5\% for IC and 11.6\% for EC).
When considering photon pairs, 
the kinematic cuts described below
depended then on the four possible photon configurations
defined previously.
\begin{figure}
 	\includegraphics[width=\linewidth]{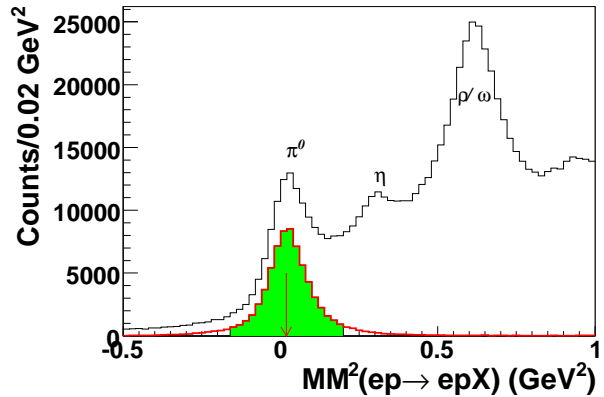}
 	\vspace*{-4mm}
 \caption{\label{fig:id}
 		(Color online)
 	   Distribution of 
 	   squared missing mass for the $ep\to epX$ reaction
	   before (black line) and after (thick red line)
	   all cuts on other variables are applied.
	   The arrow points to the pion mass, while
 	   the shaded green area correspond to the selected events.
 	   }
\end{figure}

\begin{figure}
 	\includegraphics[width=\linewidth]{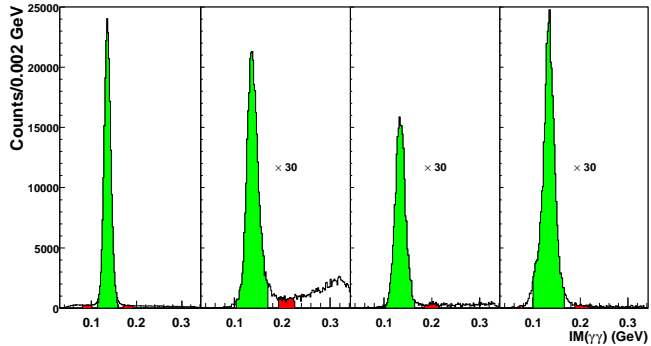}
 	\vspace*{-3mm}
 \caption{\label{fig:Mgg}
 		(Color online)
 	   Distributions of the two-photon invariant
 	   mass, after the application of all cuts
 	   described in the text, for the four
 	   configurations IC-IC, IC-EC, EC-IC, EC-EC,
 	   from left to right.
 	   The shaded areas correspond to the selected
 	   peaks (in green) and to the side-bands used
 	   in the background subtraction (in red).
 	   Note the change of scale for the last three
 	   configurations.
 	 }
\end{figure}

Events were then selected using a cut at $\pm 3\sigma$ in
the squared missing mass MM$^2(ep\to e\pi^0X)$ and a cut
in the cone angle between the expected direction of the
pion from $ep\to epX$ kinematics and the measured direction
of the two-photon system. 
This selection resulted in very clean
peaks in all kinematic correlations 
(Fig.~\ref{fig:id} gives one example)
and in the distributions 
of the two-photon invariant mass
(see Fig.~\ref{fig:Mgg}), with respectively 191k, 12k,
7k and 14k events. The small remaining background
was estimated using side-bands on the two-photon invariant mass spectra, 
for each beam helicity state
and for each of the elementary bins in ($x_B$, $Q^2$, $t$ and
$\phi$).
  
\section{$\pi^0$ asymmetry}
\label{sec:asym}
\begin{figure}
 	\includegraphics[width=\linewidth]{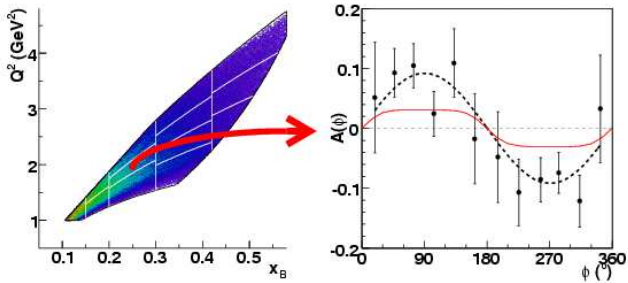}
 \caption{\label{fig:result1}
		(Color online) 
 	   Left: kinematic coverage and binning in the ($x_B$, $Q^2$) plane. 
 	   Right: $A(\phi)$ 
 	       for one of the 13 ($x_B$, $Q^2$) bins
 	       and one of the 5 bins in $t$,
 	       corresponding to
 	       $\langle x_B\rangle =0.249$, 
 	       $\langle Q^2\rangle =1.95$ GeV$^2$
 	       and $\langle t\rangle =-0.29$ GeV$^2$;
 	       the black dashed curve corresponds to a fit with
		$A\simeq\alpha\sin\phi$ 
		and the red solid curve to the JML model discussed
		in the text.
 	 }
\end{figure}

The data were divided into thirteen bins in the ($x_B$, $Q^2$) plane
(see Fig.~\ref{fig:result1}), five bins in $-t$ 
(defined by the bin limits 0.09, 0.2, 0.4, 0.6, 1 and 1.8 GeV$^2$)
and twelve 30$^\circ$ bins in $\phi$. 
The resolutions in all corresponding variables are smaller than the
bin sizes. Bin centering corrections were applied.

\begin{figure}
 	\hspace*{-6mm}\includegraphics[width=94mm]{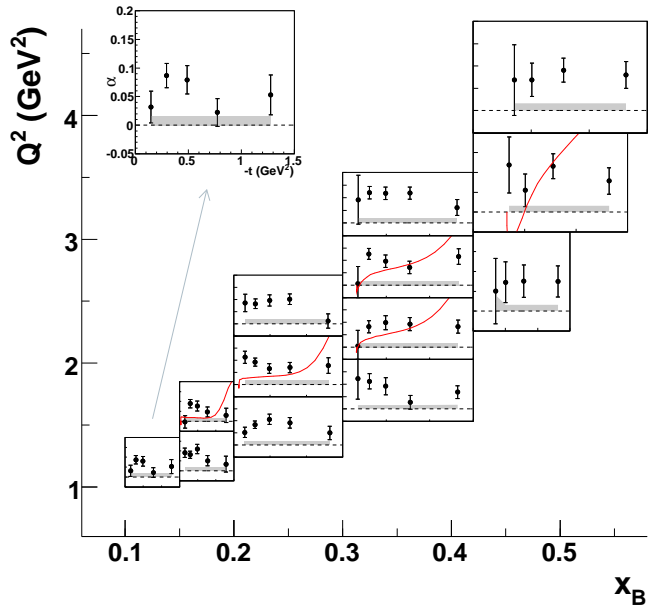}
 	\vspace*{-9mm}
 \caption{\label{fig:result2}
 		(Color online)
 	   Fit parameter $\alpha$, as extracted from $A\simeq\alpha\sin\phi$,
 	   as a function of $-t$.
 	   The location of each individual plot corresponds to the 
 	   approximate coverage in $(x_B, Q^2)$, except the upper
 	   left one (
	   an enlargement of the lower left one) which indicates
 	   the scales common to all plots.
 	   The grey areas indicate the maximal size of systematic
 	   uncertainties.
 	   For selected kinematics, the red curves correspond to the JML model 
	   discussed in the text.
 	 }
\end{figure}

Within statistical accuracy, 
the $\phi$-distributions were found to be compatible with
$A\simeq\alpha\sin\phi$ in each $t$-bin (Fig.~\ref{fig:result1}
right). The same compatibility was observed when the
$\phi$-distributions were integrated in $t$.
The determination of the asymmetry amplitude at 90$^\circ$ was
stable whether the terms in $\cos\phi$ and $\cos2\phi$ in Eq.~(\ref{eq:bsa})
were included in the fit or not.
Figure~\ref{fig:result2} gives the values
of $\alpha$ 
in the 62 ($x_B$, $Q^2$, $t$) bins considered.
By conservation of angular momentum, 
the helicity-flip transverse amplitude, and thus
$A$ and $\alpha$, are identically zero as $t$ reaches its
kinematic limit $t_0$, corresponding to $\pi^0$s emitted in the 
direction of the virtual photon. 
At small $x_B$, 
the value of $-t_0$ is smaller than our first bin limit 0.09 GeV$^2$ 
(corresponding to the proton-energy detection threshold), which is why $A$
does not go to zero.
The increase of $-t_0$ 
explains the missing $t$-bins at large $x_B$.

Systematic uncertainties arise from the
event selection, as well as from the choice of the fit function
used to extract $\alpha$. 
Together, they were estimated at 0.016. The comparison between two separate analyses 
led to increase this value for two points in Fig.~\ref{fig:result2}.
The beam polarization measurements
induce an additional overall relative uncertainty of 3.5\%. 
The data set may be found in \cite{CPDB}

\section{Discussion of the results}
\label{sec:disc}
As seen in Fig.~\ref{fig:result2}, the
measured beam spin asymmetries are systematically of the
order of 0.04 to 0.11, over a wide kinematic range in
$x_B$, $Q^2$ and $t$. In particular, there is no evidence of
a decrease of $\alpha(t)$ as a function of $Q^2$.
This is a clear sign of a non-zero
$LT^\prime$ interference among the amplitudes describing the 
$\gamma^*p\to p \pi^0$ reaction.

In the GPD formalism, only the 
longitudinal amplitude, dominant in the Bjorken regime, 
is calculated. The present
evidence of non-zero transverse terms indicates that
it may be necessary to perform a $L/T$ separation in order to isolate
the longitudinal part of the cross section. 

\begin{figure}
 	\includegraphics[width=\linewidth]{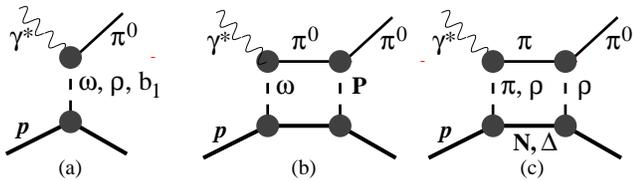}
 \caption{\label{fig:regge}
 	   Diagrams describing the neutral pion production
 	   in the JML model:
 	   (a) Pole terms.
 	   (b) Box diagram with elastic $\pi^0$ rescattering.
 	   (c) Box diagram with charge exchange ($\pi^+N, \pi^+\Delta^0,
 	   \pi^-\Delta^{++}$ are the three intermediate states considered).
	   The exchanged mesons are to be understood as the corresponding
	   Regge trajectories, and {\bf P} stands for the Pomeron.
 	 }
\end{figure}
          A Regge-type model (JML) describes the pion photo- and
electroproduction according to the diagrams in Fig.~\ref{fig:regge}.  The
model parameters are tuned to describe the photoproduction data. In
particular the strength of the $b_1$ exchange term is adjusted to
reproduce the linearly polarized photon beam asymmetry~\cite{Gui97}.
In extending the model to the case of electroproduction, vertex
electromagnetic form factors are adjusted to reproduce the DESY
data~\cite{Bra75}.  The application to the kinematic range of the
present data is then an extrapolation of the model, 
which will be fully described elsewhere~\cite{Lag07}
and reproduces the target-spin asymmetry~\cite{Che06}.  
When considering
the pole terms, only the $b_1$ exchange, through its interference with
the $\rho$ and $\omega$ exchanges (because of opposite parities), may
generate a non-zero beam spin asymmetry.  Treating the box diagrams in
the approximation of on-shell intermediate particles yields the solid
curves presented in Figs.~\ref{fig:result1} and~\ref{fig:result2}.  As
apparent in Fig.~\ref{fig:result1},
the model generates sizeable $\gamma$ and $\beta$ terms in Eq.~(\ref{eq:bsa}),
corresponding respectively to a $TT$ interference
due to the pole terms of Fig.~\ref{fig:regge}a
and to a $LT$ interference
due to the box diagrams of Fig.~\ref{fig:regge}c.

\section{Summary}
\label{sec:sum}
Sizeable beam-spin asymmetries for exclusive neutral pion electroproduction
off the proton
have been measured above the resonance region for the first time.
These non-zero asymmetries imply that both transverse and longitudinal
amplitudes participate in the process. 
The determination of the longitudinal cross section in the kinematic
regime considered here,
and the subsequent extraction of polarized generalized parton distributions,
may then necessitate to perform a $L/T$ separation.
For the same purpose, measurements at still higher values of $Q^2$
would be crucial in providing evidence for the expected decrease of the
transverse cross section.
Presently, the only available model
to calculate this observable is based on Regge theory. 
It reproduces the magnitude of the asymmetries at intermediate
values of $t$, but does not exhibit the measured kinematic dependences.
Beam-spin asymmetries for exclusive $\eta$ electroproduction,
as well as
cross sections for $\pi^0$ and $\eta$ meson production,
will be considered in forthcoming publications.

\begin{acknowledgments}
We acknowledge the outstanding efforts of the staff of the 
Accelerator and Physics Divisions at JLab,
as well as of the technical staff at DAPNIA-Saclay and IPN-Orsay,
that made this experiment possible.
This work was supported in part by 
the French Centre National de la Recherche Scientifique 
and Commissariat \`{a} l'Energie Atomique,
the U.S. Department of Energy 
and National Science Foundation,
the Italian Istituto Nazionale di Fisica Nucleare,
the Korean Science and Engineering Foundation, 
the U.K. Engineering and Physical Science Research Council. 
The Jefferson Science Associates (JSA) operates the 
Thomas Jefferson National Accelerator Facility for the United States 
Department of Energy under contract DE-AC05-06OR23177. 
\end{acknowledgments}

\bibliography{biblio} 

\end{document}